\documentclass[aps,prb,tightenlines,twocolumn,
floatfix,superscriptaddress,nofootinbib,nobibnotes]{revtex4}
\usepackage[colorlinks,allcolors=blue]{hyperref}
\usepackage{upgreek}
\usepackage{epsfig}
\usepackage{tikz}
\usepackage{graphicx}
\usepackage{epstopdf}
\usepackage{fancybox}
\usepackage{makeidx}
\usepackage{mathrsfs}
\usepackage{appendix}
\usepackage{amsmath}
\usepackage{amsfonts}
\usepackage{natbib}
\usepackage[figuresright]{rotating}
\usepackage{floatpag}
\usepackage{natbib}
\usepackage{marginnote}
\usepackage{enumerate}
\usepackage{mathtools}
\usepackage{soul}
\usepackage{color}
\usepackage{braket}

\DeclareMathAlphabet{\mathpzc}{OT1}{pzc}{m}{it}

\newcommand{\mbf}{\mathbf}

\newcommand{\bs}{\boldsymbol}
\newcommand{\bss}{\boldsymbol{\sigma}}

\begin{document}

\title{Non-equilibrium criticality \\
and efficient exploration of glassy landscapes with memory dynamics}

\author{Yan Ru Pei}
\email[Email: ]{yrpei@ucsd.edu}
\affiliation{Department of Physics, University of California, San Diego \\ La Jolla, CA 92093}
\author{Massimiliano Di Ventra}
\email[Email: ]{diventra@physics.ucsd.edu}
\affiliation{Department of Physics, University of California, San Diego \\ La Jolla, CA 92093}

\begin{abstract}
Spin glasses are notoriously difficult to study both analytically and numerically due to the presence of frustration and metastability. Their highly non-convex landscapes require collective updates to explore efficiently. Currently, most state-of-the-art algorithms rely on stochastic spin clusters to perform non-local updates, but such ``cluster algorithms" lack general efficiency. Here, we introduce a non-equilibrium approach for simulating spin glasses based on classical dynamics with memory. By simulating various classes of $3d$ spin glasses (Edwards-Anderson, partially-frustrated, and fully-frustrated models), we find that memory dynamically promotes critical spin clusters during time evolution, in a self-organizing manner. This facilitates an efficient exploration of the low-temperature phases of spin glasses.
\end{abstract}

\maketitle

%{\bf Corresponding Author: } Yan Ru Pei, phone: 8577078376, email: yrpei@ucsd.edu \\
%
%{\bf Statement of Significance: } Spin glasses are difficult to simulate due to metastability. Most algorithms rely on clusters, replicas, or multi-canonical ensembles to escape energy barriers during simulation, but they perform well only for a restricted class of models. The memory dynamics we propose in this work provides a novel approach to explore glassy landscapes, by leveraging its non-equilibrium critical properties. It makes no explicit use of clusters or replicas, and is based on forward integration of simple ODEs that couple spins to memory variables. The dynamics navigates the low temperature phase of various spin glasses with improved efficiency over state-of-the-art algorithms. This reduces the computational cost of glass simulations, and may provide insights in the role of memory in glassy dynamics. \\
%
%{\bf Preprint: } arXiv:2102.04557 \\
%
%{\bf Classification: } Physical Sciences: Physics \\
%
%{\bf Key words: } spin-glass, memory, non-equilibrium criticality
%
%\onecolumngrid
%\raggedbottom
%\pagebreak
%\twocolumngrid

\section{Introduction}

The study of spin glasses has contributed substantially to our understanding of a wide variety of phenomena \cite{barthel,rbm,birds}, much beyond the complex magnetic models for which they were first introduced~\cite{ea}. In their most basic form, these systems are described by the following simple Hamiltonian \cite{ising}:
\begin{equation}
\label{spinH}
H = - \sum_{ij} J_{ij}s_is_j,
\end{equation}
where the spins, arranged on some $d$-dimensional lattice, acquire the values, $s_i=\pm 1$, and interact via coupling constants, $J_{ij}=\pm 1$, with $\mbf{J}$ a multivariate random variable taken from some distribution. 

Despite the deceptively simple form, the energy landscape of the model Hamiltonian (\ref{spinH}) is highly non-trivial \cite{mezard_spin,replica} for most conceivable distributions of $\mbf{J}$. Decades of mathematical ingenuity have culminated in efficient (namely, polynomial-time) algorithms for computing the partition function of any realization of $\mbf{J}$ in two dimensions  \cite{onsager,kasteleyn,edmonds,barahona}, but an efficient algorithm to simulate glasses in $d>2$ remains elusive. In fact, finding the ground state of a three-dimensional glass with arbitrary bonds was shown to be NP-complete \cite{istrail}, with the task of computing its partition function shown to be NP-hard \cite{goldberg}. This hardness fundamentally limits the efficiency of any stochastic algorithm.

Earlier approaches for simulating the model Hamiltonian (\ref{spinH}) were based on sequential Metropolis updates \cite{metro,glauber}, and modern extensions of this methodology have also been proposed \cite{sa,landfill,population,walksat,eo}. However, these algorithms are generally plagued by a large dynamical critical exponent, making the simulation largely inefficient \cite{monte_intro}. 

Later on, a method based on the synchronous update of a large correlated cluster of spins was suggested \cite{fk,sw,wolff}, which proved to be very effective for the $2d$ ferromagnetic Ising model. Unfortunately, despite several modifications made to account for the non-locality of frustration \cite{nied,kbd}, these cluster algorithms still struggle for high-dimensional glasses. The main reason behind their inefficiency is the tendency for the cluster percolation process to be persistently hyper-critical, due to a mismatch of critical temperature and cluster percolation ratio \cite{kbd2,mem_clus}. This means that the largest cluster component generally covers the entire lattice \cite{bond_frus}, resulting in a trivial global spin-flip in most cases. 

At the present stage, the best known method for taming the above issues is to generate clusters based on replica overlaps (referred to as the ``isoenergetic cluster move'' (ICM) method) \cite{houd,icm}, which reverses the percolation ratio \cite{mem_clus}. This is generally coupled with replica exchange methods such as parallel tempering (PT) for efficient thermalization \cite{PT}. Unfortunately, this approach relies heavily on the dimension of the lattice, and still tends to over-percolate in certain temperature ranges. Another recent trend is to use machine learning techniques to help identify efficient clusters \cite{deep_ising}, by putting the hidden nodes on the edges (or plaquettes) of the lattice \cite{clus_bm}. In some cases, the efforts towards this direction have been halfhearted attempts in under-employing the representative power \cite{rbm_rep} of Boltzmann machines in modeling the Boltzmann distribution of the Ising glass, resulting in mathematically equivalent formulations of the traditional cluster algorithms \cite{nied}. 

Here, instead, we propose a novel non-stochastic approach to efficiently {\it learn} the critical clusters of the glass during dynamics,  without any algorithmic aid\footnote{This approach is an application of the more general computing paradigm known as {\it memcomputing}~\cite{diventra13a}, which has been successful in the solution of a variety of problems ranging from constrained optimization to unsupervised learning \cite{DMMperspective,forrest,sean_3sat,mode_assist}.}. In sharp contrast to previous stochastic methods, which treat the spins and time as discrete variables, we first linearly relax \cite{sdp} the spin variables, and then couple them to memory variables. The coupled spin-memory system is then evolved in {\it continuous} time. 

The memory variables {\it learn} from the evolution of the interacting continuous spins, and their magnitudes correspond to the (non-uniform) percolation ratios that help generate critical clusters. This, in turn, induces {\it non-local} updates of the spins, allowing them to easily transit between different Gibbs states of the glass. The evolution of the spins and memory variables occurs {\it simultaneously}, meaning that the memory variables do not ``wait'' for the spins to equilibrate before updating themselves. This process induces a {\it non-equilibrium criticality} which persists throughout the entire evolution of the system, regardless of the underlying temperature or lattice size. We verify this by simulating the memory dynamics on three different classes of $3d$ spin glasses: the Edwards-Anderson\cite{ea}, partially-frustrated, and the fully-frustrated model\cite{tiling} (EA, PF, and FF) models. Similar results for other types of spin glasses on various graph structures \cite{sk,sphere,rbm} may be reproduced using the codes associated with this work\cite{github_ising}.

\begin{figure}
\centering
\includegraphics[scale=0.55]{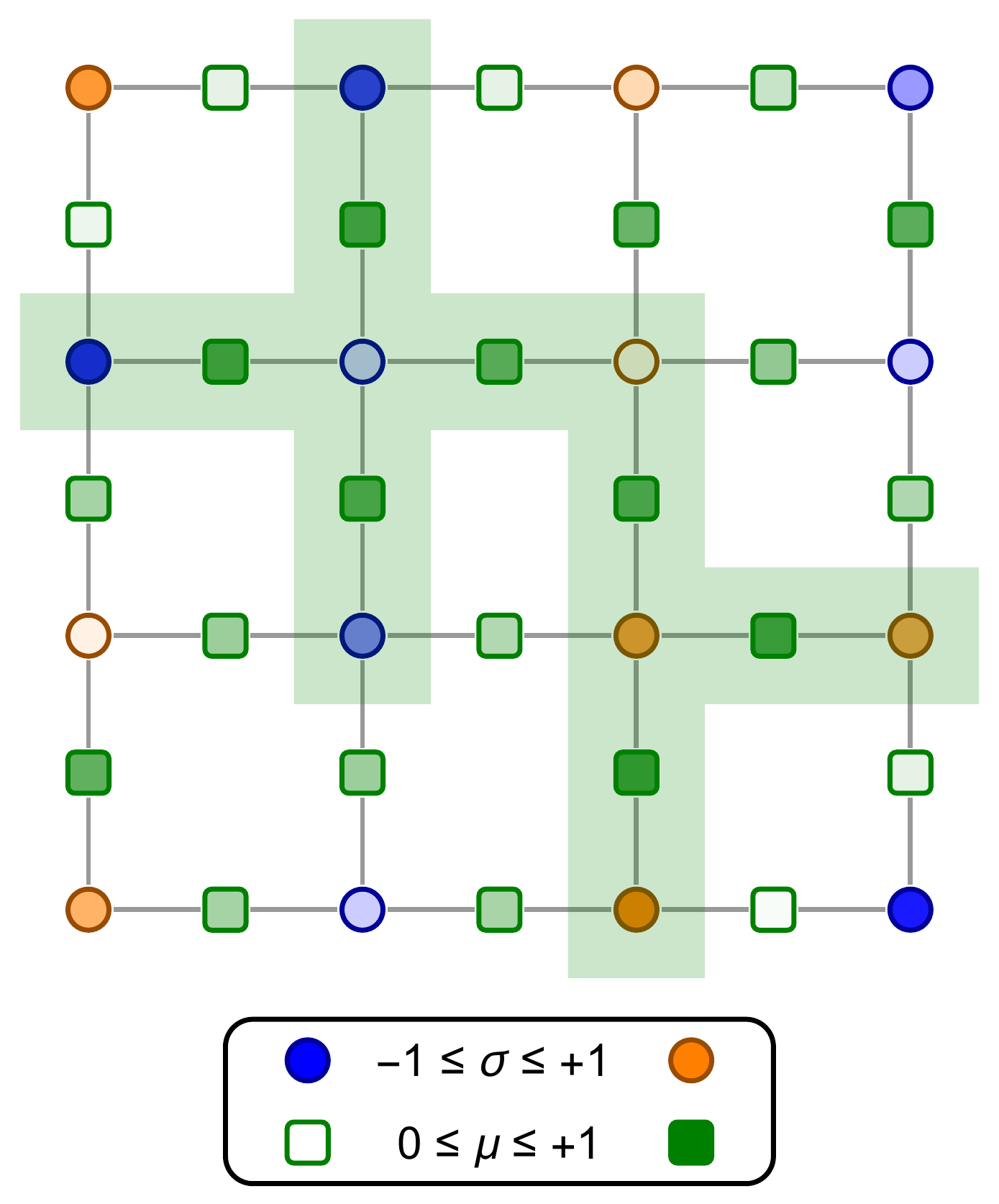}
\caption{\label{fig_mem} An instance of using memory to learn long-range dynamics in a 2d ferromagnet, where both spins and memory variables are linearly relaxed. The memory variables ``live'' on the bonds (denoted by green squares), and they are coupled to the interaction between adjacent spin states (denoted by blue/orange circles), in such a way that positive interactions increase the memory magnitudes. A potential memory cluster is shaded in green, where the memory variables can be interpreted as percolation ratios.}
\end{figure}

\section{Memory dynamics}
 
To introduce the memory dynamics, we first introduce the continuously relaxed spin glass Hamiltonian \cite{sphere}, 
\begin{equation}
\label{eq_sphere}
H = -\sum_{ij} \big( J_{ij}\sigma_i\sigma_j - \frac{1}{2}\mu_{ij}(\sigma_i^2 + \sigma_j^2) \big),\quad \sigma_{i}\in [-1,+1],
\end{equation}
where $\bs{\mu}$ are non-uniform {\it memory variables} acting as dynamic Lagrange multipliers for constraints on the spin magnitude. If $\bs{\mu}$ were fixed in time, then the standard dynamics \cite{replica}
\begin{equation}
\label{volt}
\partial_t \sigma_i = -\nabla_{\sigma_i}H = \sum_j \big( J_{ij}\sigma_j - \mu_{ij}\sigma_i \big),
\end{equation}
would suffer from long auto-correlation times (critical slowing down), due to the presence of metastable states, even when the spins are continuously relaxed.

\begin{figure*}[t!]
	\includegraphics[width=\textwidth]{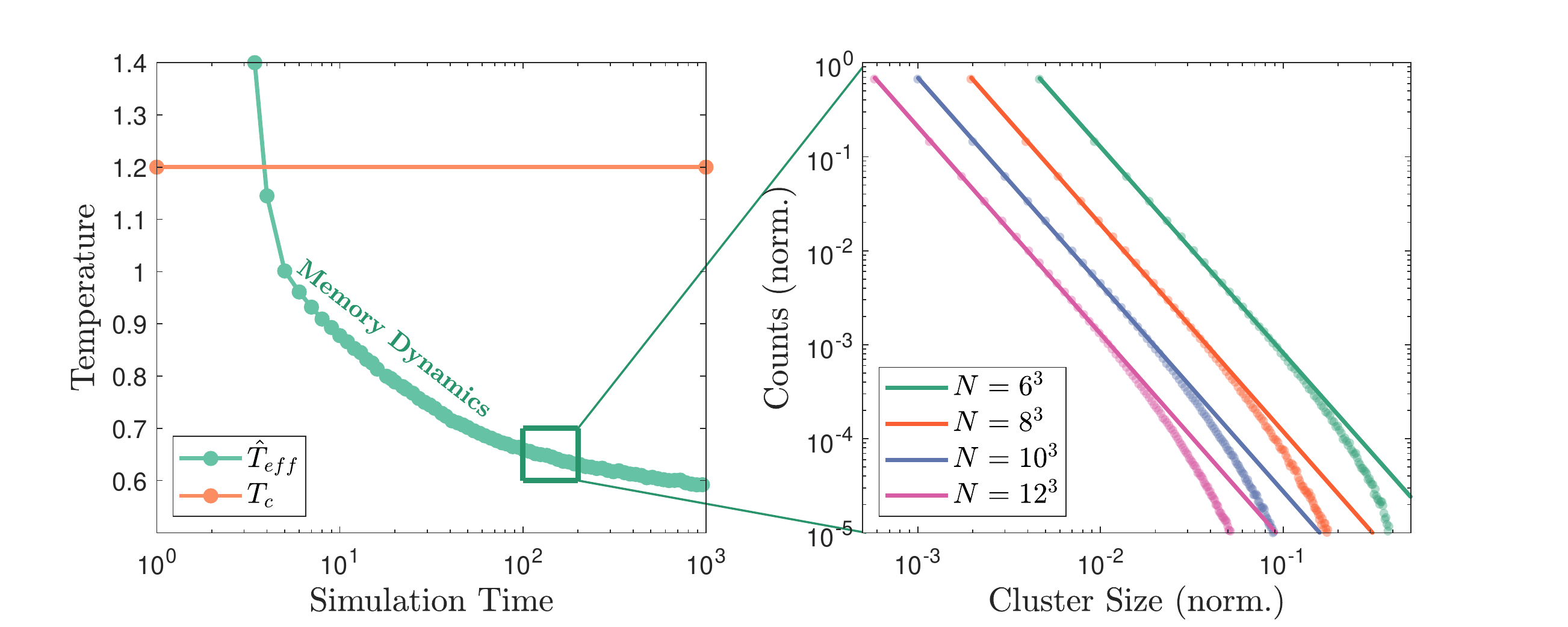}
	\caption{\label{fig_temp}
		Simulations performed on 1000 fully-frustrated $3d$ Ising glasses (see SM \ref{app::tiling}) sized $6^3$ to $12^3$. (Left) The effective temperature, $T_{eff}$, of the memory dynamics is tracked in time (see SM \ref{eff_temp} for how we estimate this temperature) for the $6^3$ lattice. The critical temperature, $T_c$, of the fully frustrated glass is determined using the crossing of Binder's cumulant (see SM \ref{crit_temp}). Note that the time it takes $T_{eff}$ to dive below $T_c$ is extremely short, less than 4 units of time, after which the memory dynamics remain persistently below $T_c$. (Right) An arbitrary point in time $t_0 = 10^2$ is chosen, and the cluster size distribution (CSD) is collected over the disorder realizations, and a $\Delta t = 2^6$ time window. Other than the tailing drop-off resulting from finite-size effects, the CSD follows a {\it power-law} decay with the Fischer exponent being $\hat{\tau} = 2.20 \pm 0.01$ for all simulated sizes. See Fig.~\ref{fig_mclus} in the SM for further empirical evidence that the Fischer exponent is insensitive to the underlying temperature and lattice size, though it is dependent on the frustration profile of the underlying glass.}
\end{figure*}

Therefore, in order to efficiently escape these states, we can {\it continuously} deform the energy local minima, and transform them into saddle points \cite{forrest,DMM2} by letting the memory variables to evolve as
\begin{equation}
\label{mem}
\partial_t \mu_{ij} = (J_{ij}\sigma_i\sigma_j - \gamma),
\end{equation}
where $\gamma$ is some constant restricting the growth of $\mbf{\mu}_{ij}$ \cite{sean_3sat}, and $\mu_{ij}\in [0,1]$. By 
providing dynamics to the variables $\mu_{ij}$, we see that the ``gradient term'' $\sum_{j} J_{ij} \sigma_j$ for the spins in 
Eq.~(\ref{volt}) is compensated by the ``cluster-like'' update term $-\sum_j \mu_{ij}\sigma_i$ in the same equation (see discussion below). 
 
By simulating the coupled Eqs.~(\ref{volt}) and~(\ref{mem}) until the system reaches a fixed time-out, we can take $s_i = \textrm{sgn}(\sigma_i)$ for a recorded state that minimizes the Ising energy in Eq.~(\ref{spinH}). Many numerical strategies can be used to improve the stability and convergence properties of the simulation \cite{sean_3sat}. They are also included in the codes of the repository \path{PeaBrane/Ising-Simulation} \cite{github_ising}, which can be used to directly reproduce Figs. \ref{fig_temp} and \ref{fig_scale}. The particular numerical implementation we used in this work is given as 
\begin{equation}
\label{mem_num}
\begin{split}
\dot{\sigma}_i &= \alpha \sum_j J_{ij}\sigma_j - 2\beta \sum_{j}x_{ij} \sigma_i \\
\dot{x}_{ij} &= \gamma C_{ij} - y_{ij} \\
\dot{y}_{ij} &= \delta x_{ij} - \zeta, 
\end{split}
\end{equation}
where $C_{ij} = \frac{1}{2}(J_{ij}\sigma_i\sigma_j + 1)\in [0,1]$, $\mbf{y}$ is a secondary {\it long-term memory} ensuring stability \cite{sean_3sat}, and $\alpha,\beta,\gamma,\delta, \zeta$ are time-scale parameters, fixed for all system sizes. We used the Euler method to integrate forward the above equations. More implementation details and the choice of parameters are given in the Supplementary Material (SM) Section \ref{mem_imp}. 
 
\section{Dynamical critical clusters} 

Before showing numerical results, we provide an understanding of why such a memory dynamics would be efficient in simulating spin glasses. First of all, we note that the memory variables should {\it not} be considered as standard dual variables \cite{lpnn}. Instead of being coupled to the spin constraints (the second term of Eq.~(\ref{eq_sphere})), the memory evolution is explicitly coupled to the state of interaction between spins, $J_{ij}\sigma_i\sigma_j$, as written in Eq.~(\ref{mem}), and this is crucial for simulating frustrated systems\footnote{The memory variables may also be defined on different unit cells \cite{kbd2,kbd3}, such as on plaquettes \cite{kbd} for the fully frustrated Ising model \cite{villain}.}. Furthermore, since we are simulating the system at {\it non-equilibrium}, we do not have to worry about using acceptance schemes \cite{mala,hmc} to tame numerical truncation errors \cite{numerical}, which do not seem to play a major role in the stability and efficiency of our dynamics \cite{mem_numerical}. 

If we bound the memory variables between $0$ and $1$ (see Section \ref{mem_imp} in the SM), we can interpret them as {\it probabilities} of forming open bonds in a {\it weighted percolation process} \cite{perc_weight}, from which critical clusters can be formed \cite{percolation}, as drawn in Fig.~\ref{fig_mem}. 

To see why the memory dynamics are {\it critical}, let us first assume that the memory variables are already at the critical percolation threshold. If one memory variable $\mu_{ij}$ is then perturbed, say, below the critical value, this effect will propagate throughout the entire lattice \cite{SOC}. In turn, this will suppress the cluster-like update term $-\sum_j \mu_{ij}\sigma_i$, making the gradient term $\sum_{j} J_{ij} \sigma_j$ relatively dominant (see Eq. (\ref{volt})). This will avalanche the Ising energy to a lower value \cite{forrest}, resulting in the sudden appearance of more satisfied interactions ($J_{ij}s_is_j > 0$). In response to these interactions, the memory variables $\mu_{ij}$ will increase until they organize to some new critical configuration (see Eq. (\ref{mem})).

To provide additional evidence of criticality, we have numerically extracted the cluster size distribution (CSD) for a fully-frustrated Ising model \cite{percolation}, as generated by the memory variables averaged over disorder at an arbitrary point in time (see Fig.~\ref{fig_temp}). While most state-of-the-art algorithms fail to generate critical clusters even at the critical temperature $T_c$ \cite{houd,icm,mem_clus}, the memory clusters are persistently critical at all temperatures (with the estimation of non-equilibrium temperature outlined in SM \ref{eff_temp}). In other words, we see that the criticality is self-organizing (SOC\cite{SOC,brain}) through the entire simulation, and it avoids critical slowing down at all points in time\footnote{Note that, it is not necessary for us to algorithmically connect the memory clusters and flip them discretely (see SM \ref{crit_perc}), because such cluster-update features are implicitly present in the equations of motion for the spin evolution (see Eq. (\ref{volt})). However, for extremely frustrated and aging glasses \cite{aging,tiling} (see SM \ref{app::tiling}), these occasional algorithmic interventions do help slightly with the relaxation time during simulations.}. It should be noted that SOC is not an intrinsic property of frustrated short-ranged spin glasses\cite{no_soc}, and this is evident in Fig.~\ref{fig_clus} of the SM, which displays a persistently hypercritical CSD for stochastic clusters generated by the Swensden-Wang (SW) and ICM rules. This also confirms the inefficiency of traditional cluster algorithms for frustrated systems\cite{kbd3,mem_clus}. 

Note that the SOC behavior of the memory induced clusters is verified for multiple $2d$ and $3d$ finite-ranged spin glasses (planted or not) with a few $3d$ examples (EA, PF, and FF models) given in Fig.~\ref{fig_mclus} of the SM. There is strong evidence that the Fisher exponent of the CSD is only dependent on the frustration ratio of the underlying glass, but independent of the effective temperature or lattice size. We encourage the readers to experiment with glasses in higher dimensions and other connectivity structures\cite{sean_3sat,mode_assist} using the available codes \cite{github_ising}.

\section{Finding a glassy ground state}
 
\begin{figure*}[t!]
	\includegraphics[width=\textwidth]{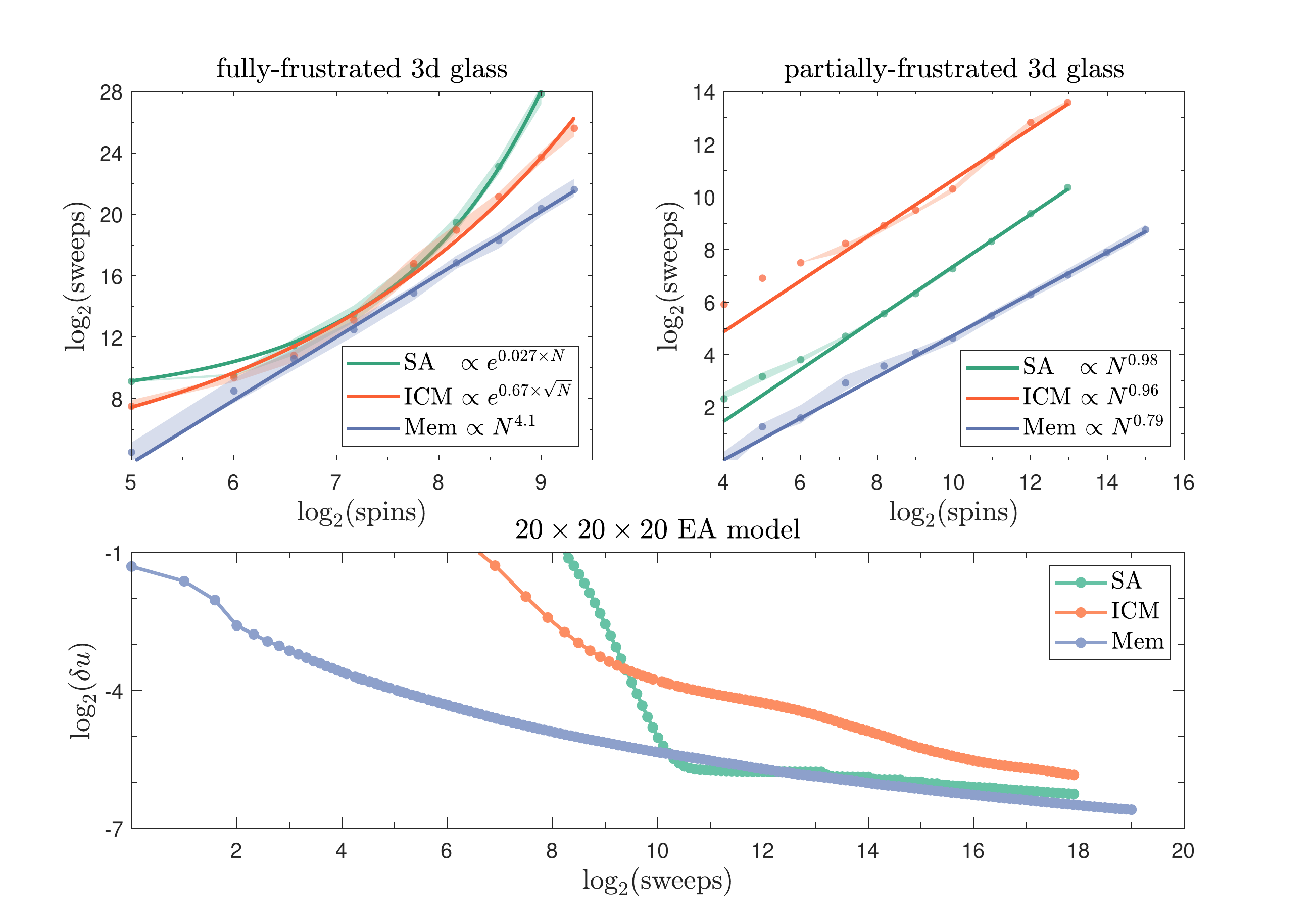}
\caption{\label{fig_scale} 
(Top) Scalability of the median number of sweeps for the fully- and partially-frustrated $3d$ spin glass (FF and PF) for simulated annealing, parallel tempering with isoenergetic cluster move, and memory dynamics (SA, ICM, and Mem), measured as the total number of sweeps (see SM \ref{comp_imp} for justification) on a log-log scale. Statistics are collected over $400$ runs, and the shaded region denotes the $40$-th to $60$-th percentile, with the fitting done with log-linear regression. The estimated scaling constants for SA, ICM, and memory are $\{ 0.027 \pm 0.0005, 0.67 \pm 0.03, 4.1 \pm 0.1 \}$ for FF and $ \{ 0.98 \pm 0.01, 0.96 \pm 0.05, 0.79 \pm 0.02 \} $ for PF, as noted in the legends. The fitting is done on the last $6$ data points for every algorithm.
(Bottom) The deviation of the best mean Ising energy from the expected ground state found over a fixed sample of 400 $3d$ Edwards-Anderson (EA) glass realizations, monitored in simulation time (sweeps). The energy returned by the memory dynamics is mostly below the other two throughout the simulation. Note that since the EA model is not planted, it cannot be verified whether the true ground state of the sample is reached (see SM \ref{ea} for further discussion).
}
\end{figure*} 
 
Finally, we show that the aforementioned non-equilibrium critical behavior allows us to find the ground state of spin glasses efficiently. First of all, we note that below the critical temperature $T_c$, the energy landscape of a spin glass becomes highly non-convex, and most algorithms fail to efficiently navigate it. For ``artificial" glassy instances where the finite residual entropy \cite{villain} (ground state degeneracy) is suppressed via the coupling of local interaction states \cite{rbm_loops,tiling}, the inefficiency of these simulations is exposed most prominently when the temperature is lowered to the $T = 0$ limit. To benchmark the efficiency of an algorithm for glass simulations, one can record the relaxation time, or equivalently the time-to-solution (TTS) to the ground state over a sample of glass realizations. We compare the memory dynamics against simulated annealing (SA) (see SM \ref{SA}) \cite{sa} and ICM (see SM \ref{ICM}) \cite{PT,houd,icm}. Note that ICM is the best known replica-based algorithm for simulating spin glasses in any dimension. See Section \ref{sto} and \ref{mem_imp} in SM for detailed discussions on how the TTS can be fairly measured for the different algorithms in terms of the number of sweeps\footnote{As discussed in SM \ref{sto} and \ref{mem_imp}, the TTS measure is made more favorable for SA and ICM so that the efficiency improvement of the memory dynamics is more convincing. There are many technical issues with directly measuring the wall-time or FLOPS, as mentioned in SM \ref{comp_imp}. The number of FLOPS can be easily extracted by multiplying the number of sweeps with the number of spins, thus adding a linear power to the time complexity measurements of all algorithms. For the readers interested in the absolute scale of wall time, solving a worst-case instance from $400$ fully-frustrated glass realizations sized $8\times 8\times 8$ on a single core with simulated annealing would take around a week.}.

With this goal in mind, we use a class of $3d$ glass instances where the frustration ratio can be controlled \cite{tiling} (see SM \ref{app::tiling} for an explanation of how these instances are generated). Since these instances are planted, the ground state energies are known in advance. This way we can verify the correctness of the algorithm. To perform the evaluation , we generate fully-frustrated $3d$ glasses \cite{tiling} up to size $8\times 8\times 10$, and partially-frustrated\cite{tiling} $3d$ glasses up to size $32\times 32\times 32$ (see SM \ref{app::tiling} for details), with $400$ randomly generated instances per size. For each size, we evaluate the efficiency of every algorithm by collecting its sweep statistics up to the $60$-th percentile (see SM \ref{app::tts}), and estimating the scaling behavior of the median number of sweeps. The implementation used for the scalability test is detailed in SM \ref{mem_imp}.

To dispel beliefs that we are fine-tuning the parameters of the memory dynamics just to solve planted benchmarks, we perform the test also on the prototypical EA model\cite{ea} for baseline reference, without changing the parameters. Since the EA model is not planted, it is exponentially hard to verify correctness of the algorithms. Instead, we use a fixed sample of $400$ random realizations of $20\times 20\times 20$ EA models for all algorithms, and monitor the log-deviation of the best Ising energies found so far (above the best known ground state energy \cite{ea_gs}) throughout the simulation (see SM \ref{ea}).

As shown in Fig.~\ref{fig_scale}, for the fully-frustrated instances, while the scaling of standard stochastic algorithms are well-fitted by super-polynomial functions (an exponential for SA and a sub-exponential for ICM), the scaling of the memory dynamics is well-fitted by a polynomial up to the maximum size we have tested. For partially-frustrated instances, all three algorithms appear to scale polynomially, with the memory dynamics having the lowest power. For the EA instances, the returned energy of the memory dynamics is mostly below the other two algorithms, and appears to continue evolving asymptotically at a lower energy as well.

In a word, not only is the memory dynamics more efficient in navigating the glassy landscapes at low temperature, but it is also more efficient in discovering ``deep" solutions. Most importantly, its efficiency has been empirically proven to be general on complex spin glasses. Nevertheless, one should note that the algorithm operates at non-equilibrium (unlike SA and ICM), so at the present stage, it is not clear how equilibrium statistics can be efficiently sampled. This is a work in progress.

\section{Conclusions}

In this work, we have introduced a new approach to simulate spin glasses based on the coupling of (linearly relaxed) spins with memory variables. We have shown numerically that the generated memory-induced spin clusters are critical, and the memory dynamics are efficient in finding the ground state of fully-frustrated Ising spin glasses in $3d$, even using the basic forward Euler discretization scheme. As a future development, the introduction of an appropriate discretization and acceptance scheme \cite{hmc} may endow the memory dynamics with the detailed-balance property \cite{metro}, making it applicable to simulating equilibrium dynamics of glasses even at finite temperature. This would allow the algorithm to be interfaced with modern stochastic algorithms, which may be useful for generating critical clusters for bosonic quantum spin and gauge systems \cite{qmc_cluster,disorder_gauge,glass_qcd}. 

Furthermore, a foreseeable generalization would be to apply this technique to simulating glasses on more general graph structures and interaction states \cite{ising_connected,potts}. Finally, it would be interesting to study the fundamental mechanism behind the criticality of memory, and its property of inducing nonlinear solitonic behavior in frustrated systems \cite{toda,crit_dynamic,nonlinear} which has been shown empirically in the past \cite{forrest}.

\section{Acknowledgments}

This material is based upon work supported by the National Science Foundation under Grant No. 2034558. Y.P. would like to thank Firas Hamze for stimulating discussions on the entropic properties of the tiling cubes, and Zheng Zhu for clarifying the implementation details of the ICM algorithm. All the numerical results presented in this study have been done on a single core of an AMD EPYC server. They can be reproduced using the codes in the repository \path{PeaBrane/Ising-Simulation} \cite{github_ising}. The repository is a complete suite for Ising Simulation in MATLAB that is maintained and developed by Y.P. 

\raggedbottom
\pagebreak
%\bibliography{SUSYref.bib}
\bibliography{crit_arxiv.bbl}
\onecolumngrid
\raggedbottom
\pagebreak

\appendix
\renewcommand\appendixname{SM}

\begin{center}
{\bf \large Supplementary Material}
\end{center}

\section{Critical Percolation}
\label{crit_perc}

Very briefly, percolation is a random process on a graph where a bond is opened (or a site is ``occupied") with a given probability $p$, and this usually generates multiple clusters on the graph (for now assume the graph is a lattice) connected by open bonds \cite{percolation}. For most graphs, there is a percolation threshold $p_c$ such that when $p < p_c$, all the clusters are finite, and when $p>p_c$, there is a unique giant cluster spanning a constant fraction of the lattice. One of the most important characterizations of the percolation process is the finite cluster size distribution (CSD), $n(s)$, which counts the number of clusters of a given size $s$ (excluding the single infinite cluster). In both the subcritical and supercritical regime ($p \neq p_c$), $n(s)$ decays exponentially with respect to $s$, while at criticality ($p = p_c$), $n(s)\sim s^{-\tau}$ decays as a power law with the critical exponent $\tau$ referred to as the {\it Fischer exponent}, which is one of the many scale-free properties of critical percolation \cite{percolation}. The majority of spin models can be translated into a modified percolation process \cite{fk}, which has been the inspiration of many cluster algorithms over the past few decades \cite{wolff,kbd,houd}. In another work we have analyzed the efficiency of these cluster algorithms both analytically and empirically \cite{mem_clus}. \\

Both the ICM algorithm (see Section \ref{ICM}) and memory dynamics (see Section \ref{mem_imp}) can be naturally studied from a percolation perspective. During the Houdayer cluster formation in the ICM algorithm \cite{houd}, spin sites with negative overlap between a replica pair can be interpreted as occupied sites, and sites with positive overlap are unoccupied sites. Randomness is introduced into the system in the form of thermalization \cite{mem_clus} generated by Metropolis sweeps and replica exchanges. There are certain procedures ensuring that the largest cluster size does not span the entire lattice. First, the cluster move only occurs between replica pairs of sufficiently low temperature, and second, whenever the number of negative sites exceeds half the spins, one of the replica is flipped globally to suppress the percolation process. However, despite these restrictions, it is shown that the algorithm still fails to be efficient in general, as it is heavily reliant on the underlying graph structure \cite{icm,mem_clus}. \\

Unlike the ICM algorithm, the percolation process defined by the memory variables we have introduced in the main text is a bond percolation process. However, the more important distinction is that the memory variables are continuous, meaning that they naturally induce a weighted graph, where each edge weight denotes the percolation probability on the bond. It has been suggested that a weighted lattice may display different critical properties than the unweighted counterparts \cite{perc_weight}. In this work, we update the distribution $n(s)$ at each simulation time unit (instead of each adaptive time step to ensure efficient and unbiased sampling \cite{thin}). We find that the distribution $n(s)$ follows a {\it power-law} decay with the giant component being absent. This suggests that the memory induced percolation process is near criticality, so that the memory dynamics are efficient in sampling the underlying glass near $T_c$ \cite{kbd1,kbd3}. \\

To make practical use of the clusters generated by the memory variables, we can perform a Swensden-Wang (SW) update \cite{sw} on these clusters as an intelligent restart method in a digital implementation of the memory dynamics. The SW update entails flipping each cluster independently with probability $\frac{1}{2}$. Note that this method satisfies detailed balance even if the percolation ratios defined by the memory variables are not uniform across the graph. Alternatively, one can also opt to perform a Wolff update\cite{wolff} which involves flipping one randomly chosen cluster, though the two methods are similar in efficiency when the CSD is critical. Note that this algorithmic step is not central to the efficiency of the memory dynamics, though it does help to increase the TTS by a small constant factor. \\

\begin{figure}
\centering
\includegraphics[width=\textwidth]{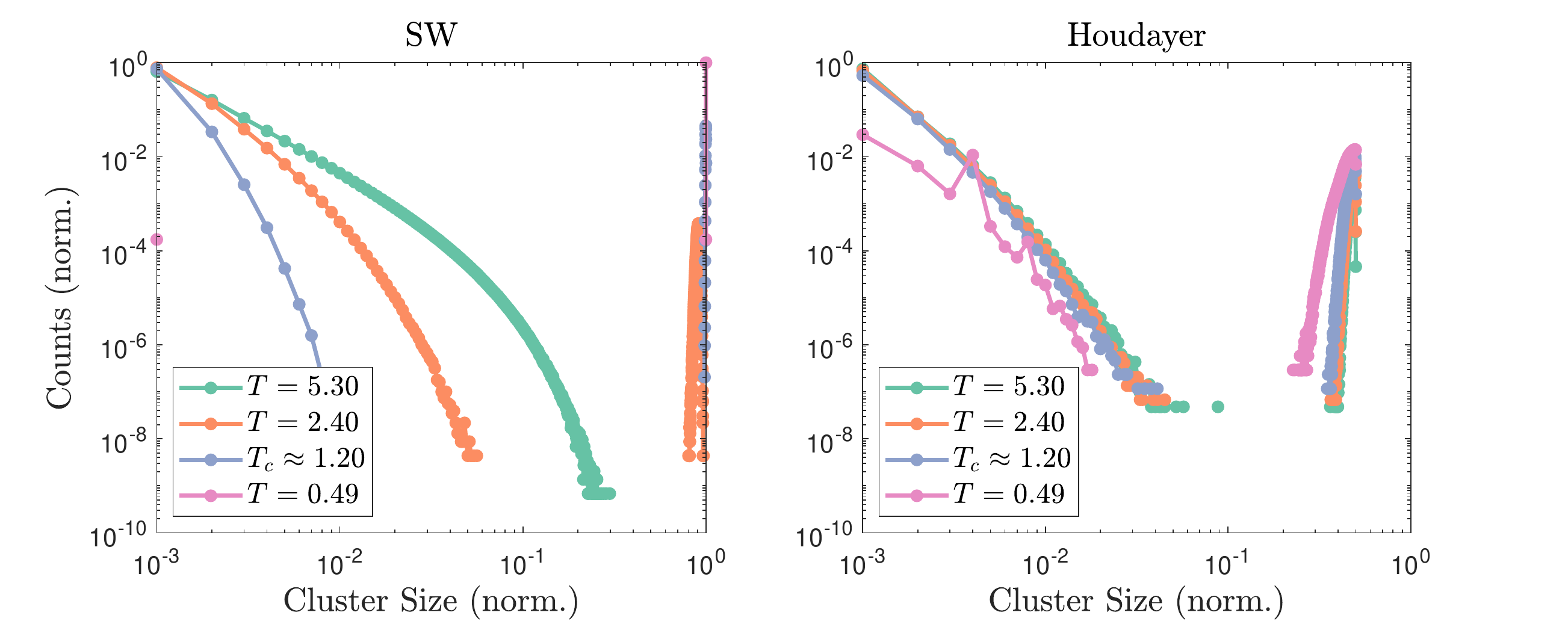}
\caption{\label{fig_clus} The cluster size distribution (CSD) for the Swensden-Wang (SW) and Houdayer percolation rules on a fully frustrated $3d$ Ising glass sized $10\times 10\times 10$. The equilibrium statistics is collected over $100$ realizations of disorder, simulated with parallel tempering (PT) over $2^{15}$ sweeps with a waiting time of $2^{19}$. Note that both classes of clusters are hypercritical at the critical temperature $T_c \approx 1.20$ (see Section \ref{crit_temp}). While the SW clusters become increasingly hypercritical as the temperature is lowered, the ICM clusters appear to be hypercritical at all temperatures, even when the largest component is restricted to half the lattice size (see Section \ref{ICM}).}
\end{figure}

\begin{figure}
\centering
\includegraphics[width=\textwidth]{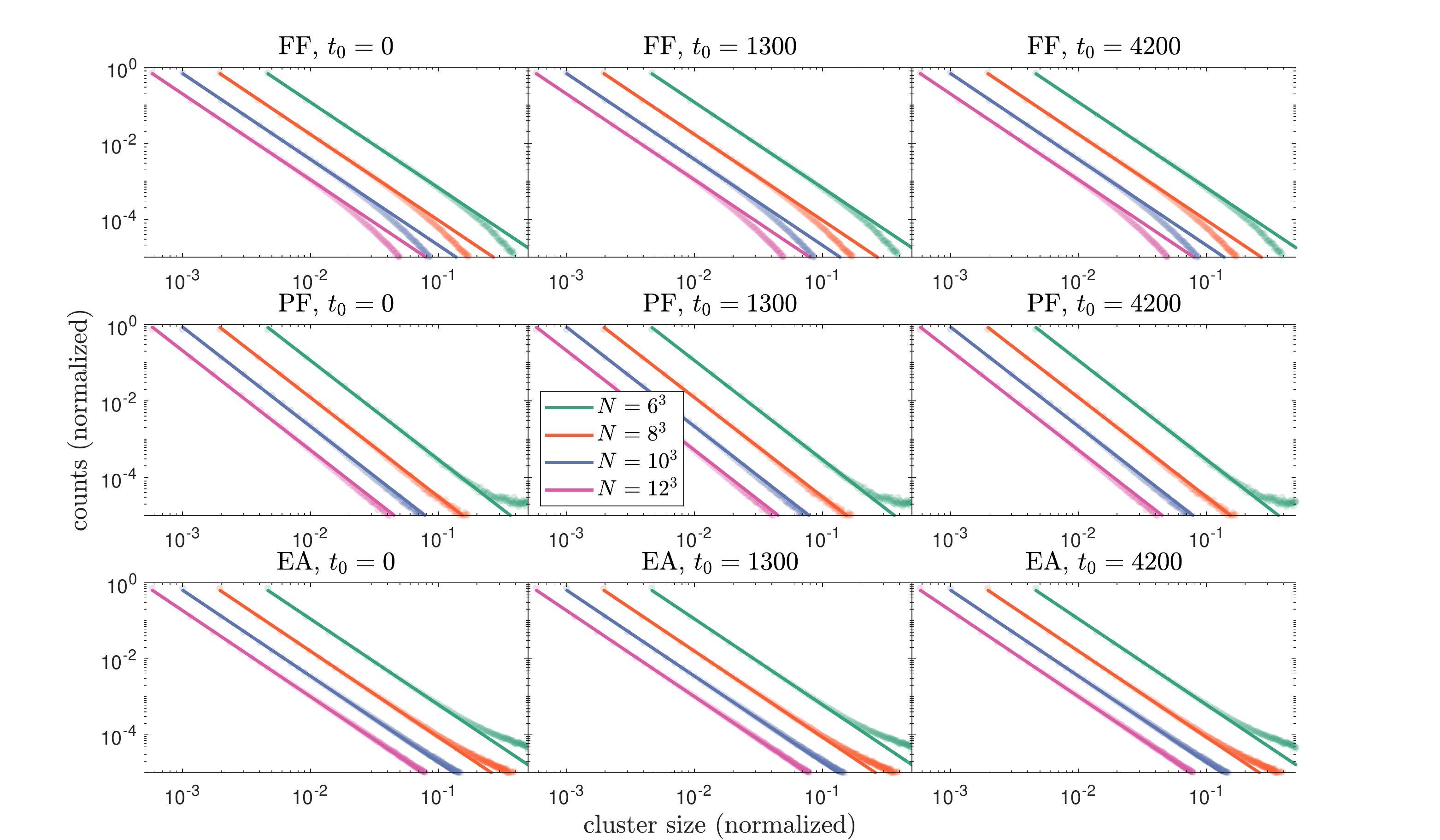}
\caption{\label{fig_mclus} The clusters generated by the memory variables appear to be near criticality at any time during evolution on the fully-frustrated, partially-frustrated, and Edwards-Anderson (FF, PF, and EA) $3d$ Ising glass, with the decay power seeming to depend only on the frustration ratio of the underlying glass (a higher frustration yields a lower power). The equilibrium statistics is collected over $400$ realizations of disorder and a time window of $\Delta t=2^6$ beginning at simulation times $t_0 = \{0, 1300, 4200\}$.}
\end{figure}

For stochastic clusters generated using algorithmic bond-formation rules, such as Swensden-Wang \cite{sw} or Houdayer \cite{houd} clusters, the CSD tends to be hyper-critical for frustrated models \cite{kbd2}, due to the mismatch between the critical temperature $T_c$ and the critical threshold for the effective (bond- or site-, respectively) percolation ratio $p_c$ \cite{kbd3}. In the SW case, this can be expressed as
\begin{equation*}
T_c \ll \frac{2}{\log\big( 1/(1-p_c) \big)},
\end{equation*} 
meaning that the CSD is already critical far above the critical temperature. Similar expressions can be derived for the Houdayer clusters as well. A study focusing on the inefficiency of stochastic clusters is detailed in another work \cite{mem_clus}, and we here simply show that the empirical CSD for SW and Houdayer rules on the fully-frustrated $3d$ Ising glass is persistently hyper-critical, as shown in Fig.~\ref{fig_clus}, meaning that cluster algorithms employing such non-local update rules cannot be generally efficient, especially if the frustration ratio of the underlying glass is large. On the other hand, the clusters generated by the memory variables are critical at any time during evolution regardless of the underlying frustration profile of the glass, though the Fisher exponent seems to depend on the frustration ratio of the underlying glass model (see Fig.~\ref{fig_mclus}).

\section{Complexity}
\label{comp_imp}

Practically speaking, the most direct measure of complexity is the wall time required to run a given algorithm until the solution is reached. However, this measure of complexity suffers from the uncertainty due to a number of irrelevant variables that are hard to measure or control, originating mainly from not only the details of the algorithmic design, but also the programming language implementation and the hardware architecture \cite{parallel}. For example (as further discussed in SM \ref{mem_imp}), the efficiency of simulating the ODEs for memory dynamics will depend on how well the code is vectorized and (further downstream) how the SIMD instructions are handled by the CPU \cite{code_vec}, none of which are directly controllable nor relevant to the time complexity of the algorithm (though they have been optimized to the extent of the authors' ability in order to push the scalability test as far as possible). \\

From a theoretical standpoint, the main focus is the time and space complexity of the algorithm, which can be roughly interpreted as the scaling of the computational cost and memory requirement of the TTS with respect to the size of the problem \cite{complexity_bible}. As the scalability is the main concern here, the actual time and memory (RAM) required to run the algorithm for a specific problem type is not of major interest, meaning that any complexity measurement and algorithmic implementation that differ in the prefactor or additional terms of smaller powers should be treated as equivalent. For example, if the time required to run an algorithm is $a N^3 + b N^2$, where $N$ is the size of the problem, then the actual prefactor $a$ and the entire second term $bN^2$ are inconsequential to the time complexity, which is $O(N^3)$. Finally, since it is clear that SA \cite{sa}, ICM \cite{icm}, and memory dynamics should scale linearly\footnote{This is because only the spin states are stored in the memory during the simulation, and a sweep is essentially a one-pass algorithm that does not create any new array structures. For replica-based algorithms like ICM, the number of replicas are constant with respect to system size (see Section \ref{ICM}), so they only contribute a pre-factor to the space complexity.} with respect to memory (RAM) requirements \cite{stress}, we will not be focusing on that here. \\

The only measure of interest should then be the time complexity of both algorithms running the same class of instances. For both algorithms, we estimate the time complexity numerically using the median statistics of TTS over $400$ tiling glass instances \cite{tiling}, with the $40$-th to $60$-th percentiles  reported to verify the robustness of the algorithm. The reason we chose to record the median instead of the mean\footnote{There is a debate on whether one should measure the median or mean for empirical studies of scalability. Practically speaking, measuring the median is much more computationally inexpensive because it requires only solving up to the $50$ percentile of the sample in terms of TTS. However, the mean TTS is of more theoretical interest, because the definition of self-reducibility is based on whether the mean complexity equates to the worst-case complexity.} is because the hardness of the instances follows roughly a log-normal distribution \cite{tiling}. One can always extrapolate the scaling of the mean TTS by doing regression on the reported median and the percentile statistics, by using a log-linear model \cite{regression}. \\

For measurement of TTS, one can measure (or estimate) the number of FLOPS (floating-point operations) required for the algorithm to find the solution. Here, we report an equivalent measure of the number of sweeps over the lattice (see Section \ref{sto} and \ref{mem_imp}). Since a sweep is an one-pass algorithm, the number of FLOPS can be evaluated by simply multiplying the number of sweeps with the system size (number of spins), which adds a linear power to the scalability. Note that this will not change whether the time complexity of an algorithm is polynomial or super-polynomial, or the efficiency comparison between different algorithms in general. For readers interested in scalability of FLOPS, we encourage them to re-implement the algorithms in their language of choice referring to our own MATLAB implementation \cite{github_ising}, which unfortunately is not streamlined for the measurement of FLOPS without sacrificing too much practical efficiency.

\section{Implementation of Stochastic Algorithms}
\label{sto}

Regardless of how efficient the employed cluster update routine is, all stochastic algorithms inevitably use an ergodic routine referred to as a {\it sweep}, where all the spins in the lattice are updated sequentially in a Metropolis-type acceptance scheme \cite{glauber,metro}. To be more precise, whenever a single spin $\sigma_i$ in the lattice is flipped (from $\sigma_i$ to $-\sigma_i$), the change in the Ising energy associated with it is
\begin{equation*}
\Delta E_i = 2 \sum_{ij} J_{ij} \sigma_i,
\end{equation*}
and the Metropolis acceptance ratio of this update is then given as
\begin{equation*}
P(\sigma_i \to -\sigma_i) = \min \Big( 1, \exp\big( -2\beta \Delta E_i \big) \Big).
\end{equation*}
A single sweep usually occurs at a constant inverse temperature $\beta$, making it generally interfaceable with a plethora of cluster or replica algorithms. \\

In most cases, it is important for the spins to be updated sequentially \cite{sequence}, as many attempts of trying to introduce synchronous update methods (such as stripe-wise updates \cite{sweep_parallel}) generally lose more in ergodicity than gain in parallel efficiency \cite{parallel}. This means that in a computational implementation, a single sweep is best limited to a single core, and it usually constitutes the most computationally intensive routine when used in conjunction with cluster updates. The complexity is simply $O(N)$, where $N$ is the number of spins, noting that the coordination number of the $3d$ lattice is fixed at $z=6$. Several modern algorithms that efficiently utilize the sweep routine include simulated annealing (SA), parallel tempering (PT), and isoenergetic cluster moves (ICM), with their scalabilities of the median TTS for the fully frustrated $3d$ Ising glass \cite{tiling} shown in Fig.~\ref{fig_scale} of the main text (with the PT and ICM algorithms combined).

\subsection{Simulated Annealing}
\label{SA}

The Simulated Annealing (SA) algorithm \cite{sa} is inspired by the physical annealing process in metallurgy, where the metal is gradually cooled from a high temperature to a low one to improve its ductility. In the context of optimization, this means that we begin with a high temperature, and gradually lower the temperature to a very small value, and this process somewhat aids the algorithm in navigating the non-convex cost function of the optimization problem to find the global minimum \cite{mezard}. This algorithm saw great success in many industrial optimization problems, such as the traveling salesman problem (TSP) \cite{tsp}. Many state-of-the-art algorithms are based on the same underlying concept, with added entropic routines for intelligent exploration of the state space \cite{population,gene}. Here, we are using the algorithm in its original form, with the annealing parameters extensively optimized. \\

In most cases, a single run of an SA routine is not sufficient to find the ground state, and more often than not, a restart routine is generally required \cite{walksat,restart}, to give the algorithms multiple chances at tackling the problem. Although there have been studies on using informed restart methods to interface with the algorithm \cite{bb}, these methods are not general, and usually only provide a pre-factorial improvement over a random restart routine. Therefore, in this work, we will simply use the random restart routine to minimize complications. \\

The three important parameters for SA are $\beta_{\min}$, $\beta_{\max}$, and $t_{\text{sweep}}$, together referred to as the {\it annealing routine}. $\beta_{\min}$ is the starting inverse temperature of the sweep, $\beta_{\max}$ is the ending inverse temperature of the sweep, and $t_{\text{sweep}}$ is the number of sweeps going from $\beta_{\min}$ to $\beta_{\max}$. Based on seminal work \cite{sa_scale}, and extensive optimization studies done previously \cite{forrest,rbm_loops}, we find that the best annealing routine of $\beta$ is linear from $\beta_{\min}=0.1$ to $\beta_{\max}=\log(N)$, where $N$ is the number of spins. Furthermore, we find that the optimal number of sweeps between restarts is $t_{\text{sweep}} = N$ regardless of the underlying frustration profile of the glass, which balances between having a sufficiently gradual annealing ratio and sufficient restart opportunities. The annealing routine can be expressed succinctly as
\begin{equation*}
\beta = 0.1 + \big(\log(N)-0.1 \big) \frac{t-1}{t_{\text{sweep}}},
\end{equation*}
where the inverse temperature is increased from $0.1$ to $\log(N)$ over $N$ sweeps.

\subsection{Parallel Tempering}
\label{PT}

As mentioned in the previous section, the major problem with using SA is the lack of a generally intelligent restart method, so in practice, most practitioners simply use a random restart routine, where all the spins in the lattice are uniformly sampled from $\sigma_i = \pm 1$. As a substantial improvement, parallel tempering (PT) replicates the lattice into multiple copies, and simulates the replicas under different temperatures \cite{PT} (usually by sweeping the lattice), and two replicas of neighboring temperatures are exchanged depending on an external rule to improve efficiency of exploring the Gibbs measure. The acceptance ratio of the exchange is given by
\begin{equation*}
P( \bss^{a} \leftrightarrow \bss^{b}) = \min\Big( 1, \exp\big( (\beta^a-\beta^b)(E(\bss^a)-E(\bss^b)) \big) \Big),
\end{equation*}
which keeps the joint distribution of the entire replicated system stationary. Intuitively, the replica at the highest temperature is essentially sampling from the uniform spin measure, and this ``random restart" propagates down the replica chain through the exchange interactions, where multiple replicas essentially ``mediate" the restart routine from $\beta_{\min}$ to $\beta_{\max}$. This is the reason why PT is often times considered as an algorithm with an implicit restart routine that is ``intelligent''. This is arguably the most general algorithm designed to work for many classes of optimization problems on different underlying graph 
structures \cite{tiling,rbm_loops,chimera}, 
including many industrial problems \cite{pt_tsp,pt_rbm}. Combined with intra- or inter-replica cluster algorithms \cite{wolff,kbd,houd}, this results in incredibly efficient stochastic algorithms, where non-local updates are complemented by an intelligent restart method. We will discuss one such algorithm in the next section. In this work, we use $n_r = 30$ replicas spaced geometrically in inverse temperature from $\beta_{\min}=0.1$ to $\beta_{\max}=\log(N)$, or
\begin{equation*}
\beta_i = \beta_{\min} \Big( \frac{\beta_{\max}}{\beta_{\min}} \Big)^{(i-1)/(n_r-1)},
\end{equation*}
based partially on existing work \cite{forrest,tiling} and our own optimization attempts.  Note that the exchange update is trivial in computational cost, as it simply involves computing an acceptance ratio and exchanging the indices of two replicas.

\subsection{Isoenergetic Cluster Moves}
\label{ICM}

The ICM algorithm is currently the state-of-the-art cluster algorithm \cite{icm} for simulating spin glasses that combines the method of PT \cite{PT} and Houdayer cluster updates \cite{houd}. In the main text, this is then used as the most representative of stochastic algorithms to compare against the memory dynamics in the TTS scaling behavior on the fully-frustrated $3d$ Ising model \cite{tiling}. A comprehensive description and the pseudocode for the algorithm is already given in the literature \cite{icm}, and also implemented in MATLAB \cite{github_ising}. Here, we provide a brief overview of the algorithm and list the parameters that we used to perform the simulations. In addition, we pinpoint the most computationally intensive routine, whose scalability we use as the time complexity of the algorithm. \\

In this algorithm, the parallel tempering algorithm \cite{PT} is coupled with Houdayer cluster moves \cite{houd}. To begin, a number of replica pairs are initialized, with the pairs spaced geometrically in temperature. After one sweep in every replica, an Houdayer update is performed for every replica pair. This cluster update flips a non-trivial cluster of spins with negative overlap between two replicas to increase the mixing rate. Finally, the parallel tempering routine attempts to exchange the temperature between two random replicas of neighboring temperatures, to improve the thermalization process. This algorithm can be used both as a sampler and an optimizer. For the latter case, one simply has to record and return the lowest energy that is sampled by the algorithm. We use the same parameters as given in Section \ref{PT}, meaning that the total number of replicas is $2n_r = 60$. \\

Note that in the modern implementation of the ICM algorithm \cite{icm}, there is an extra algorithmic step that performs a global spin flip on a replica in each pair, whenever the number of negative overlap sites exceeds half the lattice size, so that the largest cluster component never exceeds half of the lattice. Although the intention of this procedure is to suppress the percolation process through restricting the size of the giant component, similar to the intention of plaquette-based bond-formation rules \cite{kbd}, it does not fundamentally address the issue of mismatching the critical temperature and percolation threshold (as shown in Fig.~\ref{fig_clus} in the SM), meaning that Houdayer clusters are still hyper-critical despite algorithmic interventions. It is also important to note that the global spin-flip routine breaks detailed balance, because the reverse transition probability is zero, meaning that the global flip should in theory never be accepted. Further discussion of this phenomenon is given in another work \cite{mem_clus}. \\

To analyze the computational complexity of the different routines to inform a fair TTS measure, we note that the PT routine is trivial in cost (as noted in Section \ref{PT}), and in the worst-case, the Houdayer move performs either a breadth-first or depth-first search (BFS or DFS) \cite{graph} to identify the spin clusters, whose time complexity is $O(N)$ for the lattice graph. This is also the time complexity of one Metropolis sweep over a replica. Therefore, the complexity can be measured as the total number of sweeps summed over all the replicas, with the Houdayer update cost ``generously" ignored/absorbed into the total complexity as it is of the same complexity power. 

\section{Integration of Memory Dynamics}
\label{mem_imp}

Recall that the memory dynamics is formulated in continuous time, and originally conceived to be implemented on physical circuits \cite{DMMperspective}. However, it was later discovered that a carefully designed memory system is robust against noise and perturbations \footnote{In the context of our study, robustness means that multiple trajectories emanating from any initial point in the phase space will go to the optimum. Therefore, it is not necessary for us to accurately integrate any particular one of them. It is possible for us to end up in another ``desirable" trajectory after deviation from the original one, and still find the optimum in the end.}, and can be readily simulated numerically on a digital computer using basic integration schemes such as forward Euler \cite{mem_numerical}. Such a basic implementation has been proven to perform exceptionally well for multiple problem structures \cite{mode_assist,forrest,sean_3sat}, as long as the relative timescale of the memory variables (with respect to the spins) is appropriate. Of course, this leaves room for many improvements in numerical methods to speed up the rate of convergence of the dynamics when the goal is to implement the memory dynamics digitally as a practical solver. Here, we present a few improvements that we found relevant to our work. First of all, we rewrite Eq.~(\ref{mem_num}) in the main text for convenience of the reader, where $\{\alpha,\beta,\gamma,\delta,\zeta\} = \{ 0.80, 1.04, 1.67, 7.07, 2.20 \}$ are constants chosen for the $3d$ cubic graph, and are fixed for system sizes and the type of glass (as long as $|J_{ij}|=1$).
\begin{equation}
\label{mem_num_2}
\begin{split}
\dot{\sigma}_i &= \alpha \sum_j J_{ij}\sigma_j - 2\beta \sum_{j}x_{ij} \sigma_i \\
\dot{x}_{ij} &= \gamma C_{ij} - y_{ij} \\
\dot{y}_{ij} &= \delta x_{ij} - \zeta, \\
\text{where} \quad C_{ij} &= \frac{1}{2}(J_{ij}\sigma_i\sigma_j + 1)\in [0,1].
\end{split}
\end{equation}
The function $C_{ij}(\bss)$ is referred to as the {\it clause function} in the field of constrained optimization \cite{zoltan_maxsat,sean_3sat}, evaluating to $+1$ if the spin interaction is satisfied, and $0$ otherwise. Note that we use $C_{ij}$ purely to notationally interface with the literature, and the offset from $J_{ij}\sigma_i\sigma_j$ can be easily accounted for by a constant shift in the initialization of $\mbf{y}$. This set of equations and parameters is used to perform the simulation of the CSD as presented in Fig.~\ref{fig_temp} in the main text.\\

Typically, to ensure positivity of $\mbf{x}$, we would opt for an exponential growth rate given as $\dot{x}_{ij} = (\gamma C_{ij}-y_{ij})x_{ij}$, which in fact already outperforms stochastic algorithms in TTS. Nevertheless, we choose to make the growth of $\mbf{x}$ linear for faster dynamics \cite{relu}, and to guarantee that no ``hidden exponential'' is present during dynamics. However, this comes at the price of potential negativity of $\mbf{x}$ and the introduction of unstable modes. To avoid this, we can dynamically anneal the decay rate via the extra (long-term) memory variable $\mbf{y}$ \cite{sean_3sat} coupled bond-wise to $\mbf{x}$. Intuitively, the new memory variable $\mbf{y}$ grows/decays along with $\mbf{x}$ with some time lag to ensure that the relative change in the magnitude of $\mbf{x}$ is never too large nor too small. Though not necessary in most cases, positivity of these memory variables can be simply enforced by introducing explicit bounding values at each time step as such, 
\begin{equation*}
\begin{split}
x_{ij,n+1} &= \min\{\max\{ \, x_{ij,n} + dt\,(\gamma C_{ij}-y_{ij}) \, ,0\},1\} \\
y_{ij,n+1} &= \min\{\max\{ \, y_{ij,n} + dt\,(\delta x_{ij}-\zeta) \quad ,1\},10\}, \\
\end{split}
\end{equation*}
In most cases, the variable $\mbf{y}$ is only relevant to the initial transient dynamics in its purpose of suppressing the highly oscillatory memory modes\cite{dmc, circuit_analysis} (when the spin glass is far from equilibrium, a rapid relaxation of the spins induces large fluctuations in the magnitude of $\mbf{x}$). When the system relaxes slightly, the variable $\mbf{y}$ will decay quickly to its lower bound (if the parameter $\zeta$ is chosen appropriately), and will effectively serve as a constant decay of $-1$ for $\mbf{x}$. If the memory dynamics dive quickly below $T_c$ (see Fig.~\ref{fig_temp} in the main text), then $\mbf{y}$ is usually not needed, but we include it here for the sake of generality. A formal analysis of this discretization/bounding in the context of stability, absence of periodic orbits and chaos is given in the supplementary material of our previous work \cite{sean_3sat}. Note that we do not leverage any stochasticity in taming discretization errors \cite{hmc,mala}, which is further empirical proof of robustness. \\

Beside the memory variables, the step size itself can be also made adaptive \cite{adaptive_step} to further improve stability and speed up convergence. We adapt the step size as such,
\begin{equation*}
dt_{n} = \min\Big\{\max\Big\{ \, \frac{1}{\max_i(|\dot{\sigma}_{i,n}|)} \, ,2^{-5}\Big\},2^{-3}\Big\},
\end{equation*}
to regularize the maximum voltage change at every step\footnote{Note that this time step adaptive schedule we choose to use is rather unconventional. The standard adaptive schedule is to geometrically tune the step size based on the local error estimate for the purpose of efficiently simulating the solution trajectory. However, in our case, we do not require such accuracy, and employing such procedure actually decreases the rate of convergence to the optimum.}. Although not necessary, an intelligent restart method \cite{restart} can also be implemented to ensure that the energy landscape is being thoroughly explored by the memory dynamics. After initialization, the dynamics are integrated up to time $t_0 = 2^6$, and if the optimum is not found in the time duration, then we perform a SW update on clusters generated by $\mbf{x}$ (see Section \ref{crit_perc}), and continue to run the dynamics for another iteration, until the ground state is discovered or the total timeout is reached. \\

Note that the time complexity of performing an integration step is $O(N)$, which is the same as a single sweep in stochastic algorithms (see Section \ref{sto}). However, when implemented in hardware, integrating a time-step is always much faster than a sweep, because the spin updates in our integration scheme are synchronous (as with standard explicit methods for simulating multivariate ODEs), meaning that the machine code can be vectorized to interface with a single instruction, multiple data (SIMD) hardware structure \cite{code_vec}, whereas it would be incredibly difficult to do so with a sweep, even if the spin data are stored bit-wise. Again, to be generous, we do not consider such hardware overhead in TTS measures, and simply measure the TTS as the total simulation time. Note that since the adaptive time step is bounded below by a constant $dt_{\min} = 2^{-5}$, there is no cost associated with the inverse scaling of step size with respect to $N$ \cite{zoltan_maxsat}. In the end, the redundant scaling $O(N)$ is factorized out for both the TTS measures for stochastic algorithms and the integration of memory dynamics, as it only affects the polynomial order of the TTS scaling, but it does {\it not} affect whether the scaling is polynomial or exponential. \\

Furthermore, we note that the parameters $\{\alpha,\beta,\gamma,\delta,\zeta\}$ for the memory dynamics were tuned very minimally using simplex descent \cite{simplex}, with the cost function being the mean Ising energy returned for $40$ runs of memory dynamics on the $6\times 6\times 6$ uniform random-bond Ising glass \cite{nishimori}. This ensures that we are {\it not} taking advantage of the specific structure of the tiling instances by over-fitting the parameters specifically for the fully frustrated $3d$ Ising glass. This is similar to the technique used in machine learning for tuning hyperparameters such as the learning rates \cite{hyper_simplex}, where to avoid over-fitting, the parameters are tuned on some given neural network (NN) for a subset of tasks that it is designed for and then the NN is tested on another disjoint subset of tasks for cross validation of performance \cite{cross}. On the other hand, the parameters for SA and ICM (annealing schedule and temperature spacing respectively) are tuned extensively based on previous works \cite{icm,forrest}, which effectively tilts the playing field in favor of stochastic algorithms. 

\subsection{Non-local Extensions}
\label{KBD}

The main culprit behind the inefficiency of local cluster algorithms is that the bond-formation rules are defined edge-wise, such as the SW rule, so they completely ignore the non-local effects of frustration. Therefore, such algorithms are prone to over-percolate \cite{kbd1,kbd3,mem_clus}, and are generally inefficient for frustrated systems. This has inspired multiple extensions of the original SW rule to more non-local unit cells. For instance, the fully frustrated Ising model (FFIM) in $2d$ \cite{villain} can be deconstructed into unit cells of checkered plaquettes, with each plaquette guaranteeing exactly one negative interaction at the ground state. This led to the realization that bond-formation decisions can be made plaquette wise, resulting in the prototypical KBD plaquette rules \cite{kbd}. This in turn inspired a plethora of other non-local cluster algorithms \cite{kbd3,qmc_cluster} designed for other classical and quantum glasses. The KBD plaquette rule has proven to be efficient in reducing the auto-correlation time in simulating magnetization properties \cite{kbd2}, though its efficiency is still largely restricted to the $2d$ FFIM \cite{mem_clus}. \\

The equations for the memory dynamics (see Eq.~(\ref{mem_num_2})) can also be modified to interface with such non-local bond-formation rules. For instance, if we take the $2d$ FFIM, and index the checkered plaquettes (or checkered cubes in 3D) as $\square$, we can restrict the long-term memory variables to these plaquettes, and couple it to the short-term memory $\mbf{x}$ as, e.g., 
\begin{equation*}
y_{\square} = \delta \sum_{(i,j)\in \square} x_{ij} - \zeta.
\end{equation*}
Instead of simply regularizing $\mbf{x}$ as discussed in Section \ref{mem_imp}, the long-term memory now provides additional non-local information to the original edge coupled system of $\{\mbf{\sigma},\mbf{x}\}$. To make the analogy with machine learning, we note that the use of multiple layered network structures is common for pattern recognition in deep learning \cite{deep}, and here, $\mbf{y}$ acts as an additional layer of nodes, learning to recognize plaquette frustration patterns. To take this analogy further, we could introduce more layers of memory to learn successively non-local cell patterns \cite{kbd2,kbd3}, as long as stability is still guaranteed. For this implementation, we use a slightly different set of parameters $\{ \alpha, \beta, \gamma, \delta, \zeta \} = \{ 1.42,1.81,1.24,0.21,2.9 \}$, which again we find by minimal tuning. \\

\begin{figure}
	\centering
	\includegraphics[scale=0.6]{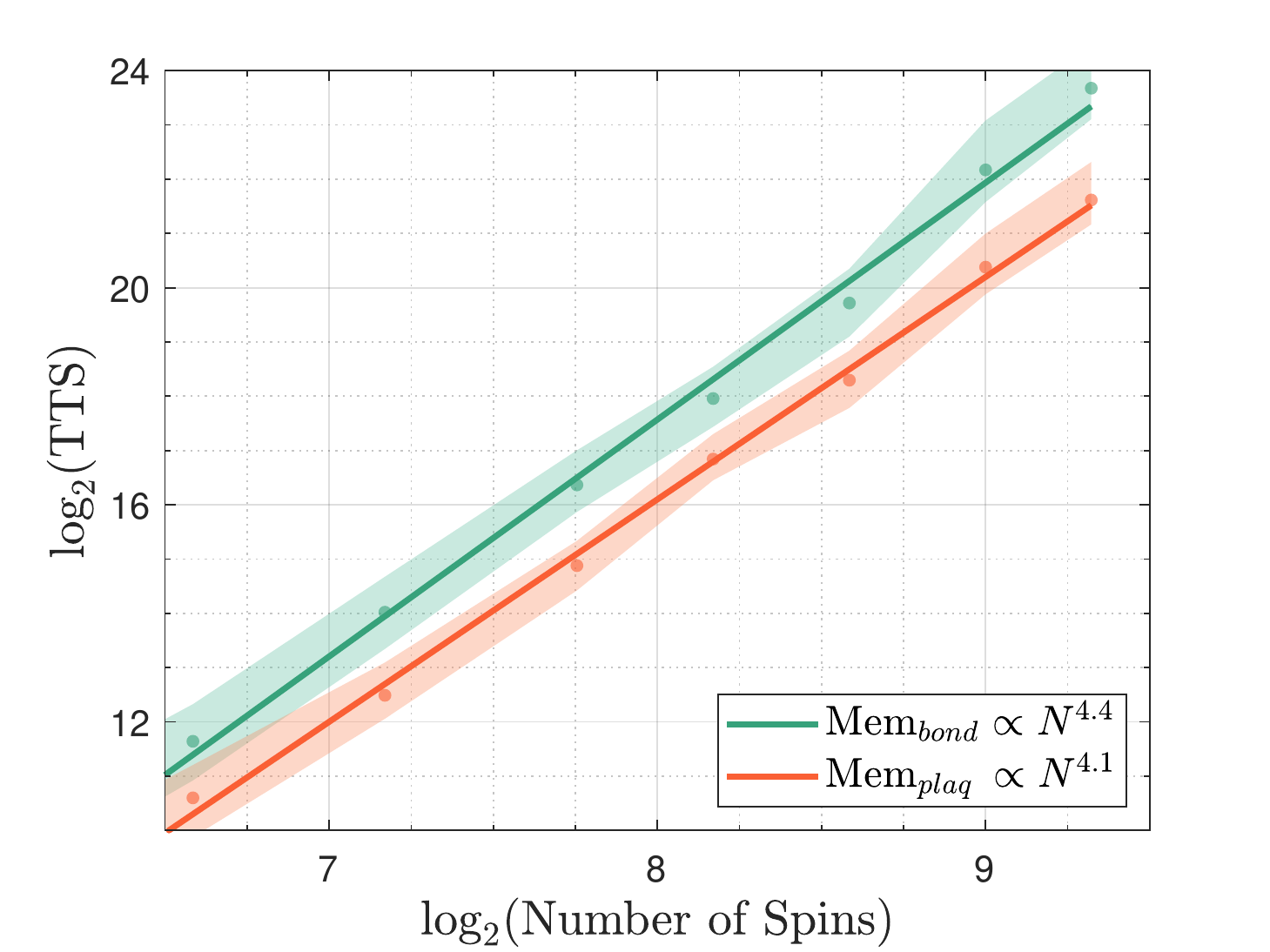}
	\caption{\label{fig_plaq} The green line is the fitted scalability for the median TTS for the bond-based long term memory (LTM) on the fully-frustrated $3d$ glass, and the orange line is for the checkered LTM on the same instances. The scaling powers are $4.4 \pm 0.1$ and $4.1 \pm 0.1$ respectively, with the checkered based LTM displaying slightly favorable power. Note that the medium TTS's are well-fitted by polynomial functions for both schemes.}
\end{figure}

For the scalability of TTS for the fully-frustrated glass as shown in the upperleft subplot of Fig.~\ref{fig_scale} in the main text, we use an implementation that assigns $\square$ to be the checkered cubes, as an extension of the KBD algorithm in $2d$. To ensure that we are not exploiting the tiling instances \cite{tiling}, we intentionally offset the checkered pattern for the long-term memory variables with the one used to generate the instances. Note that even if the checkered patterns for the tiling glasses and the long-term memory were the same, it would by no means represent an advantage. This is because the problem of finding a ground state for a tiling instance still appears to be NP-complete \cite{jigsaw}, even if the knowledge of the checkered pattern (or the planting pattern \cite{plant_ksat}) is given. In fact, even for $2d$ glasses, there is not a single cluster algorithm that is efficient even when the checkered pattern is known \cite{kbd,kbd2,tiling_hard,mem_clus}. It is interesting to note that the scalability of both the bond and plaquette memory dynamics show similar scaling powers, as shown in Fig.~\ref{fig_plaq}. 

\section{Edwards-Anderson model}
\label{ea}

The Edwards-Anderson model \cite{ea} is the ``prototypical" spin glass where the couplings $\mbf{J}$ are sampled uniformly from $\{-1, +1\}$. It is believed that the lower critical dimension of this model is $d_l = 2.5$ \cite{ea_lower}, meaning that it exhibits a non-trivial phase transition in $3d$ (even though this has not yet been mathematically proven \cite{disorder}). Therefore, the $3d$ EA model serves as a perfect ``natural" benchmark for simulation algorithms. Note that since the model does not assume a planted ground state, extracting the time-to-solution (TTS) measure (see Section \ref{app::tts}) is infeasible (because it requires actually solving for the ground state which may be exponentially hard to verify). Therefore, we opt to simply monitor the evolution of the lowest Ising energy found by the algorithm so far as a function of time (see Section \ref{sto} and \ref{mem_imp} for the measurement of simulation time) at a fixed size of $20\times 20\times 20$. \\

Even though the EA model is not planted, the EA model is translationally invariant (or rather the disorder measure is), the system is self-averaging \cite{disorder}, meaning that the ground state energy in the thermodynamic limit equals the mean ground state energies over the distribution of disorder, or
\begin{equation*}
\lim_{N \to \infty} E_{gs}(\mbf{J}) = \overline{E_{gs}(\mbf{J})},
\end{equation*} 
where $E_{gs}(\mbf{J})$ is the ground state energy of a particular realization $\mbf{J}$. This value is extracted to be $-1.7875(2)$ (from simulations performed on $3d$ EA models sized from $L=6$ to $L=14$) normalized with the number of spins \cite{ea_gs}, so we can monitor the best Ising energies found by an algorithm so far over a sample of disorder realizations with the following metric
\begin{equation*}
\log(\delta u(t)) = \log \big( 1.7875 + \frac{\overline{ E_{best} (\mbf{J}, t) }}{N} \big),
\end{equation*}
where $E_{best}$ is the best energy found so far in time $t$ for the realization $\mbf{J}$. The overline in this context denotes both the average over the glass samples and the expected operation of the algorithm (if it is stochastic). Note that this metric should decrease monotonously in time, with a smaller value meaning that the algorithm is closer to the ground state energy. For our study, we randomly generated $400$ EA samples of size $L = 20$. \\

It is important to note that this measure may become unreliable for measuring the absolute efficiency of an algorithm at low temperature, for small sample and lattice sizes, in which case the statistics of the glass sample itself (plus finite-size effects) becomes prominent. For instance, when an algorithm plateaus at some energy above the expected ground state, it is unclear whether this deviation may actually be due to the inability of the algorithm to find the ground state rather than the sample itself having a higher ground state energy than expected. And in rare cases where the ground state energy of the sample is lower than expected, the measure $\log(\delta u(t))$ then becomes undefined. Nevertheless, for sufficiently large sample and lattice sizes, it is still expected that the inefficiency of the algorithms will dominate the statistical deviations of the sample, making the logarithmic measure accurate in comparing the efficiency of the algorithms.

\section{Planted Ising Spin-glasses}

Planted Ising spin-glasses is a class of Ising instances generated by assuming a specific ground state, without sacrificing the glassy property of having a highly non-convex energy landscapes. This makes them ideal benchmarks for evaluating algorithms that simulate critical spin-glass dynamics, or finding the ground state of glass realizations. In addition, it is generally beneficial to have control over the hardness of the planted instances to offer different degrees of evaluation, and an important way of realizing this control is to design the class to offer tunable frustration ratio \cite{lattice_loops, rbm_loops, tiling}. \\

We would also like to address a common belief that the so-called ``planted" glass models are not ``real'' spin glasses. The identification of what constitutes a ``real spin glass'' is meaningless, as formally speaking, any spin-glass structure is essentially a distribution of couplings $\mbf{J}$ on some underlying graph structure \cite{fragile}, and practically speaking, there is no reason to expect planted models to be less ``physical" than traditional glass models. In fact, multiple ``planted" structures are known to exhibit glass-like behaviors \cite{barthel,qhid}, and even some highly frustrated deterministic models are glassy in nature \cite{determine1}. In another work, we show several interesting extremal properties of planted glasses in $2d$ \cite{mem_clus}.\\

A full discussion of the computational hardness of planted problems is beyond the scope of this work, but we refer the reader to \cite{spectral_sat,plant_ksat} for a formal analysis on the computational hardness of planted problems. Generally speaking, it is unknown whether we can have a method of randomly generating problem instances in the NP-complete class (planted or not) that are self-reducible \cite{pcp}, meaning that it is unclear we can generate hard\footnote{``Hard" in the sense that the problem is NP-complete. If an NP-complete problem is non-adaptively random self-reducible, then the polynomial hierarchy collapses to $\Sigma_3$. The problem of finding the ground state of planted instances is clearly not NP-complete, so it is possible that they are self-reducible (without any severe implication).} instances with the average complexity equal to the worst-case complexity. Nevertheless, empirical evidence has suggested that the fully-frustrated glass is extremely difficult for most state-of-the-art algorithms to simulate or solve \cite{tiling,tiling_hard}.

\subsection{Time-to-solution}
\label{app::tts}

Generally speaking, there are two major ways to evaluate the efficiency of an optimizer/solver in its ability to discover a ground state of a glassy instance. The first way is to allocate the solver a certain amount of time, and allow the solver to run until timeout. This evaluation method is generally done for incomplete solvers \cite{mse2019}, where the goal is to test the capability of the solver to reach the lowest energy possible within a given time. In most cases, this method of evaluation does not provide an accurate measure of the efficiency of the solver or the complexity of the problem class, and is generally biased towards greedy or local solvers. To see why, most complex systems (such as the Ising spin glass) admit a rough energy landscape \cite{replica,tiling} with an abundance of local minima (metastable states) which can be readily accessed by a greedy algorithm with random initial conditions convoluted by random noise \cite{walksat}. However, the transition from a metastable state to the global optimum (Ising ground state) requires exponential cost in time \cite{pcp}, and, in addition, requires careful coordinated non-local updates \cite{wolff,kbd,icm,forrest}. \\

A greedy solver may reach a metastable state relatively quickly \cite{3sat_phase}, but it may never reach the global optimum. On the other hand, a solver with collective dynamics may sacrifice some time to carefully establish long-range connections \cite{forrest}, and eventually reach the global optimum after being allocated sufficient time. Therefore, a more faithful measurement of the efficiency of a solver is to record the time it takes for the solver to return the global optimum (or reach a certain gap above the optimum). This evaluation is commonly known as the time-to-solution (TTS) evaluation \cite{lattice_loops}. However, to actually perform this measurement in practice, one has to know in advance what the optimum is. A way to achieve this goal is to assume (or plant) a solution in advance, and generate instances such that the optimum can be easily extracted by the generator but exponentially hard for the solver to find \cite{plant_ksat}. These {\it planted} instances can then be used to evaluate the performance of the solver, and also check the correctness of the solver by comparing its solution to the planted one.

\subsection{The Tiling Glass}
\label{app::tiling}

The tiling glass \cite{tiling} is a class of Ising spin glasses with a planted ground state energy. It is a class where the expected local residual entropy of the checkered cubes governs exponentially the hardness of the instances \cite{tiling,replica_ksat}. This type of ``planted'' glass structure has rather rich equilibrium and non-equilibrium dynamics \cite{tiling,tiling_hard,mem_clus}, making it ideal for analyzing the efficiency of spin-glass simulation algorithms. Getting back on track with the discussion on the tiling glass, we first note that the tiling glass allows for the generation of a ``fully-frustrated" spin glass where every face of the $3d$ lattice is frustrated, meaning that one cannot simultaneously satisfy all interactions in any give face \cite{signed}. More generally speaking, the frustration profile (or the sign) of 5 out of the 6 faces of a cube can be independently assigned. Therefore, one can also opt to frustrate only a portion of the 6 faces to decrease the frustration ratio (thus the expected hardness) of the planted instances. All the possible frustration profiles of a cube are enumerated in the original work detailing the tiling glass \cite{tiling}.  In Fig.~\ref{fig_scale} of the main text, the ``fully-frustrated" glass refers to the tiling glass assembled with $F_6$ cubes (where all 6 faces frustrated), and the ``partially-frustrated" glass refers to the tiling glass assembled with $F_{24}$ cubes (where 4 out of 6 cubes are frustrated)\cite{tiling}. \\

Glossing over some caveats with the problem of defining full frustration in a $3d$ lattice \cite{mem_clus}, we note that a fully frustrated lattice under the tiling construction also attains the maximal local ground state degeneracy cube-wise, and this has been demonstrated numerically to generate an extremely rough energy landscape. It should be noted that the generation of a fully frustrated hypercube with extremal frustration is highly non-trivial in $d\geq 4$ dimensions \cite{fully_frustrated}, so we limit our studies to the $3d$ glasses. To assemble a fully-frustrated $3d$ glass from fully-frustrated cubes, the cubes can be rotated randomly and assembled in a checkered pattern to introduce further disorder. In general, the local frustration ratios of the cubes, fully- or partially-frustrated, is preserved when they are assembled to the global construction, meaning that the ground state energy of the glass is simply the sum of the local ground state energies of the cubes, which can be computed in linear time. Note that for state-of-the-art solvers \cite{icm,population}, a fully frustrated construction already becomes computationally prohibitive to simulate in the periodic lattice sized $8\times 8\times 8$. 

\subsection{Effective Temperature Estimation}
\label{eff_temp}

\begin{figure}
\centering
\includegraphics[scale=0.7]{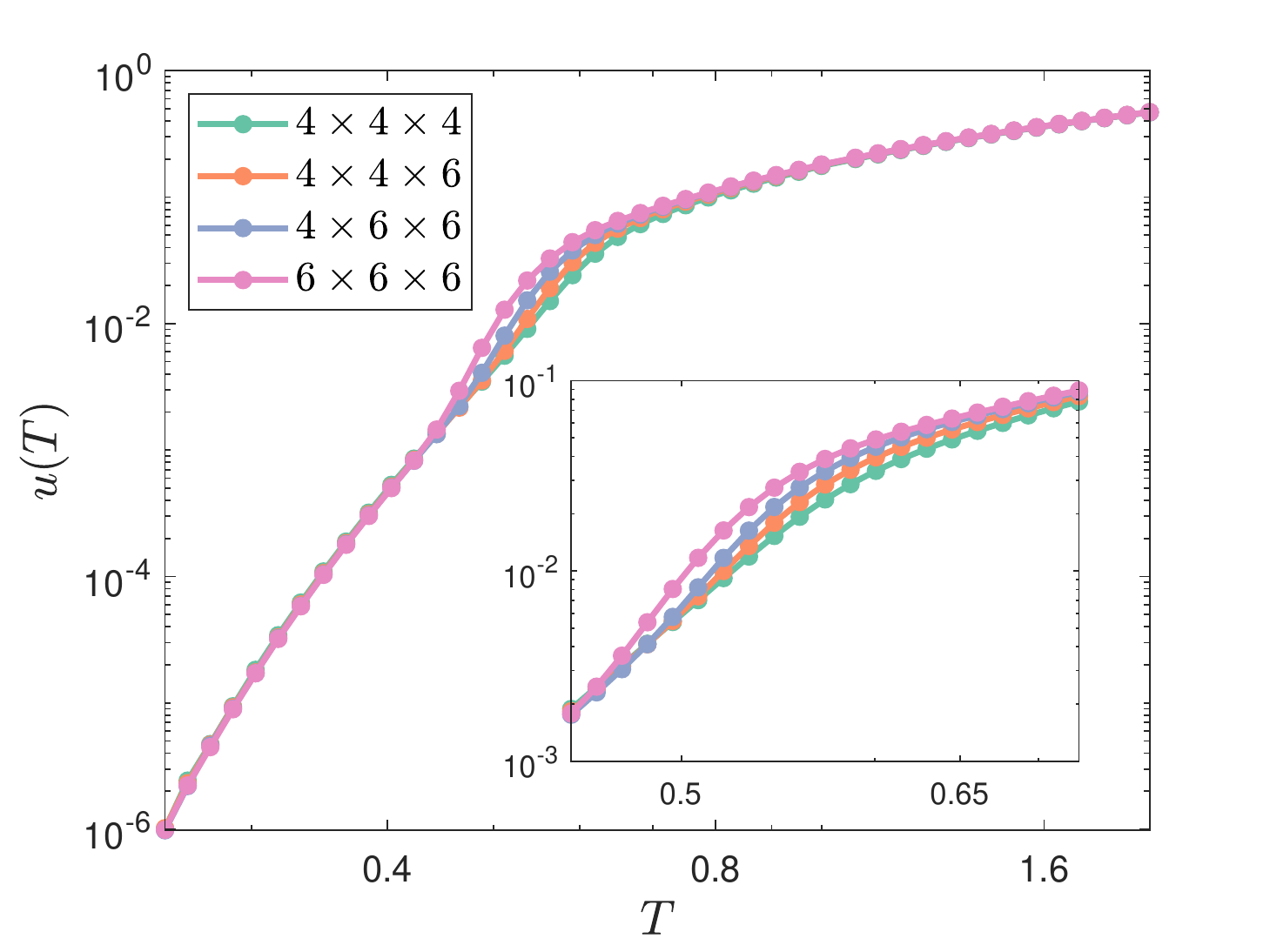}
\caption{\label{fig_ut} 
The estimated curves for the offseted intensive internal energy $u(T)$ as given in Eq.~(\ref{ut}), for fully frustrated $3d$ Ising glasses sized from $N=4^3$ to $N=6^3$. The curves slightly separate below the critical temperature $T_c \approx 1.2$ (as zoomed in on the lower right) before merging again in the low temperature phase, possibly due to non-averaging finite size effects induced by metastability in the glass phase (see Fig.~\ref{fig_bind}). The simulation is performed over $400$ disorder realizations with PT for $2^{21}$ sweeps and replicas spaced geometrically in temperature.}
\end{figure}

\begin{figure}
\centering
\includegraphics[width=\textwidth]{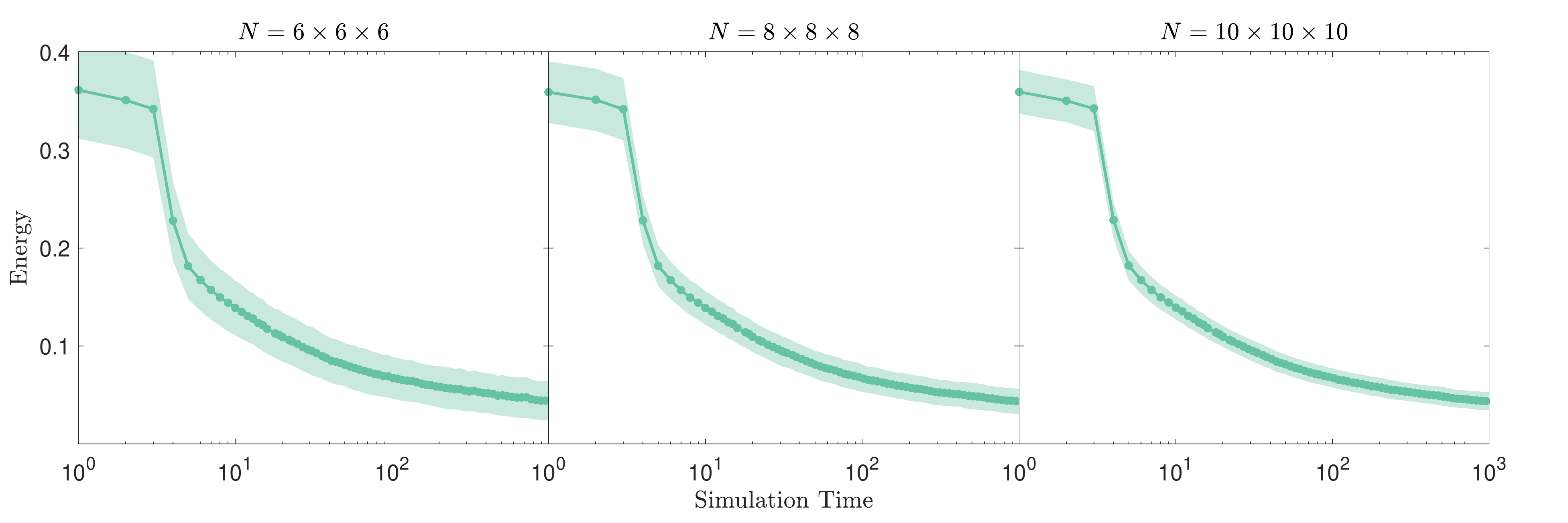}
\caption{\label{fig_evo} 
The change of internal energy in time when the fully frustrated $3d$ Ising glass is evolved under memory dynamics. Statistics are gathered over $400$ disorder realizations. The solid line represents the mean of the energy samples $\overline{u}$, and the shaded region represents the standard deviation of the samples $\hat{\sigma}_{u}$. Of particular importance is the observation that the energy variation decreases as the system size is increased, which is in line with the self-averaging property of the equilibrium internal energy, as expressed in Eq.~(\ref{self_avg}). Informally, this means that the energy variation is in fact a thermal property, instead of being induced by the random initialization of memory dynamics, which implies that the effective temperature curve shown in the left panel of Fig.~\ref{fig_temp} in the main text is in fact thermally robust.}
\end{figure}

\begin{figure}
\centering
\includegraphics[scale=0.6]{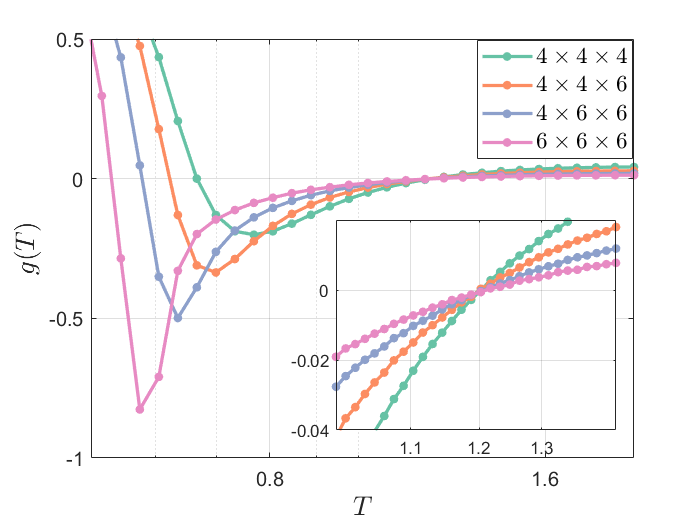} \\
\includegraphics[scale=0.45]{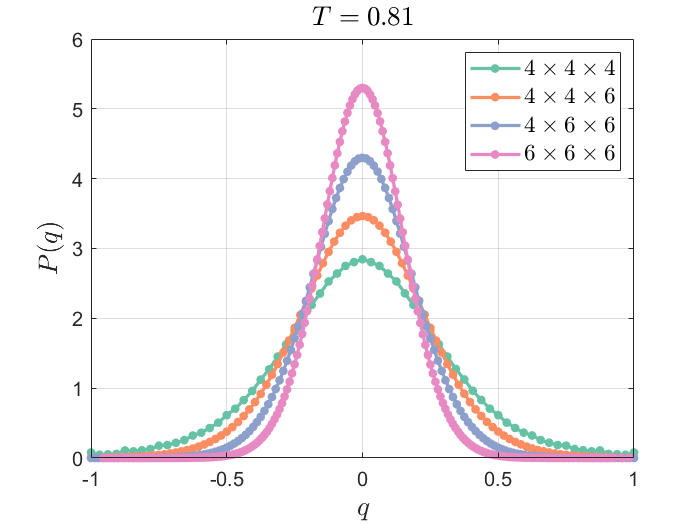}
\includegraphics[scale=0.45]{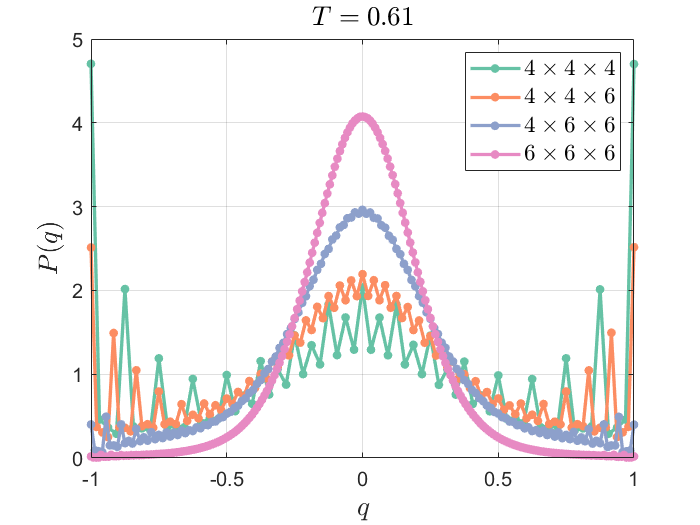}
\caption{\label{fig_bind} The curves show the Binder's cumulant $g(T)$ for spin overlap $q$ with respect to temperature (see Eq.~(\ref{eq_bind})), for fully frustrated (FF) $3d$ Ising glasses sized from $N=4^3$ to $N=6^3$. The curves intersect at $T_c \approx 1.2$, which we shall interpret as the critical temperature of the $3d$ FF glass. Interestingly, $g(T)$ at the intersection point appears to be near $0$, seeming to imply that replica symmetry is somehow ``restored", and the glass is fully disordered. This is a very interesting and uncommon phenomenon, as generally $g(T)$ is positive at the intersection point \cite{fragile}. Furthermore, we see that $g(T)$ displays a pronounced dip below $T_c$, at varying values of $T$ depending on the system size, in a range roughly corresponding to the separation of energies as seen in Fig.~\ref{fig_ut}. Again, this is possibly due to non-averaging finite size effects induced by metastability in the glass phase. We will make no attempt to investigate this phenomenon \cite{bind_anomaly}, and simply show the empirical overlap distribution $P(q)$ at the low temperature phase, noting that this distribution becomes increasingly erratic as the temperature is lowered and the system size becomes smaller. The simulation is performed over $400$ glass instances with ICM for $2^{21}$ sweeps with the replica pairs spaced geometrically in temperature. }
\end{figure}

Even though the memory dynamics operate at non-equilibrium, in the sense that the effective temperature decreases monotonously in time (see Fig. \ref{fig_temp} in the main text), the instantaneous distribution of internal energies over disorder and uniform initialization of the equations of motion does seem to converge to the Boltzmann distribution. This is reminiscent of the operation of simulated annealing (SA) \cite{sa}, though there are major differences, particularly in that SA relies on a carefully tuned annealing process approaching $T_c$ \cite{sa_scale}, while the transient memory dynamics dive below $T_c$ right away without sacrificing long-range order. \\

To see this `static' equilibrium property more clearly, in Fig. \ref{fig_evo} we observe that, for the memory dynamics, the relative variation of internal energies over disorder decreases as the system size is increased, which is an expected property of an equilibrated spin glass as the internal energy is proven to be self-averaging \cite{self-average}, or
\begin{equation}
\label{self_avg}
\lim_{N\to\infty} \frac{U_J(T)}{\overline{U}(T)} \overset{p}{\to} 1,
\end{equation}
meaning that we can extract the effective temperature of the memory dynamics at any point of time from the sample mean of the recorded energies. This follows from the general procedure of maximum likelihood estimation (MLE) \cite{regression,temp}, where the following likelihood function is to be optimized over $\beta$,
\begin{equation}
\label{LL}
\sum_{\bss \in \mathcal{S}} P(\bss) \log \big( \frac{e^{-\beta E(\bss)}}{\mathcal{Z}(\beta)} \big),
\end{equation}
where $\mathcal{S}$ is the set of spin state samples. Note that we are ignoring the disorder distribution here for the aforementioned reasons. Setting the derivative of (\ref{LL}) to zero gives us
\begin{equation*}
\begin{split}
& \partial_{\beta} \Big( \sum_{\bss \in \mathcal{S}} P(\bss) \log \big( \frac{-\beta E(\mbf{s})}{\mathcal{Z}(\beta)} \big) \Big) \\
=& \sum_{\bss \in \mathcal{S}} P(\bss) \big( -\beta E(\bss) - \partial_{\beta} \log(\mathcal{Z}) \big) \\
=& \beta\sum_{\bss \in \mathcal{S}} P(\bss) \big( U(\beta) - E(\bss) \big) \\
=& \frac{1}{T} (U(T) - \overline{E}) = 0,
\end{split}
\end{equation*}
where $\overline{E}$ is the sample mean of the energies. The temperature estimator is then given by
\begin{equation*}
\hat{T} = U^{-1}(\overline{E}),
\end{equation*} 
where again we are making no distinction between $U_J$ and $\overline{U}$. Note that $\hat{T}$ and $\hat{\beta}$ can be interchanged by virtue of the functional invariance of MLE estimators, and furthermore, the estimator $\hat{T}$ is efficient \cite{rao}. \\

For non-trivial spin glasses \cite{replica}, it is likely that the analytic form of $U(T)$ is not available, so we resort to using Monte Carlo methods to estimate it. For the tiling cubes \cite{tiling}, we can define the intensive internal energy offset by the planted ground state energy $E_0$ as
\begin{equation*}
\label{ut}
u(T) = \frac{U(T)-E_0}{N},
\end{equation*}
where $N$ is the number of spins. 

\subsection{Critical Temperature Estimation}
\label{crit_temp}

For deterministic models (or sufficiently structured glasses), one can generally look at the behavior of the spin correlation \cite{villain} or the aging profile \cite{aging} of the overlap autocorrelation time to determine the critical temperature $T_c$. However, the local rotations of the tiling construction gives rise to large variations in these order parameters over different instance realizations (even in the absence of any gauge transformation \cite{signed,rbm_loops,mem_clus}). Therefore, we will resort to using the Binder's cumulant \cite{binder} as a much more robust order parameter, which is related to the kurtosis of the overlap distribution over both the Boltzmann measure and the disorder of $\mbf{J}$,
\begin{equation}
\label{eq_bind}
g = \frac{1}{2}\Big( 3 - \frac{\overline{\braket{q^4}}}{\overline{\braket{q^2}}^2} \Big),
\end{equation}
where $\braket{\cdot}$ denotes the Boltzmann average, the overline denotes the quenched average over disorder, and $q$ is the spin overlap \cite{dynamic_overlap} defined as
\begin{equation*}
q = \frac{1}{N} \sum_i \sigma_i^{\alpha} \sigma_i^{\beta},
\end{equation*}
with $\alpha$ and $\beta$ denoting two independent replicas. In most cases, the temperature at which the $g(T)$ curves intersect for different system sizes can be taken numerically as the critical temperature $T_c$ \cite{short} (see Fig. \ref{fig_bind}). \\

There may be a disagreement on whether a crossing at the cumulant of $0$ actually accurately pinpoints the critical temperature of the fully-frustrated glass. This discussion is an open and important one, but far beyond the scope of this work. We will attempt to offer our side of this discussion in another work \cite{mem_clus}. Note that however, the exact value of the critical temperature $T_c$ is not central to the results here, which is to show the non-equilibrium critical property of memory, and prove that it can be leveraged to explore the low-temperature phase of spin glasses efficiently. In the context of this work, the critical temperature is only important for identifying the ``transient" stage of memory evolution (see left subplot of Fig.~\ref{fig_temp} in main text), which may be slightly prolonged if $T_c$ is in fact lower.

\raggedbottom
\pagebreak

\end{document}